\documentclass[12pt]{article}
\usepackage{fleqn,ccc,amssymb}
\usepackage{mathptm}
\usepackage{graphicx}
\usepackage{bm}

\title{\bf Binding Energies and Scattering Observables
in the $^\mathbf{3}$He$^\mathbf{4}$He$_\mathbf{2}$ Atomic System\thanks{This
work was supported by the Deu\-t\-s\-che For\-s\-ch\-ungs\-ge\-mein\-schaft
(DFG), the Academia Sinica, the NSC of Taiwan (ROC), and the Russian
Foundation for Basic Research}}

\author{W. Sandhas\address{PI Universit\"at Bonn,
Endenicher Allee 11-13, D-53115 Bonn, Germany},
E. A. Kolganova\address{BLTP, JINR, 141980 Dubna, Moscow Region,
Russia}$^{\rm c}$,
A. K. Motovilov$^{\rm b}$, and
Y. K. Ho\address{IAMS, Academia Sinica, PO Box 23-166, Taipei,
Taiwan 10764, R.\,O.\,C.}}

\begin{document}
\thispagestyle{empty}

\maketitle

\begin{abstract}
\small The  $^3$He$^4$He$_2$ three-atomic system is studied on the
basis of a hard-core version of the Faddeev differential
equations. The binding energy of the $^3$He$^4$He$_2$ trimer,
scattering phase shifts, and the scattering length of a $^3$He
atom off a $^4$He dimer are calculated using the LM2M2 and TTY
He-He interatomic potentials.
\end{abstract}
\bigskip

\medskip

The system of three He atoms is of great interest for various questions as,
e.g., the formation of liquid helium drops, superfluidity or Bose-Einstein
condensation. The literature on experimental and theoretical studies of the
symmetric $^4$He three-atomic system is rather large (see, e.\,g.,
\cite{GrebToeVil}--\,\cite{Bressani} and references cited therein). In
contrast, the asymmetric $^3$He$^4$He$_2$ system has found comparatively
little attention. To the best of our knowledge the numerical studies of the
$^3$He$^4$He$_2$ trimers were performed only in Refs.
\cite{EsryLinGreene}--\,\cite{Roudnev}. Except Ref. \cite{Roudnev}, there are
no scattering calculations reported for this system.

In  \cite{KMS-JPB}, \cite{MSSK} the symmetric $^4$He three-atomic system was
studied on the basis of a mathematically rigorous hard-core version of the
Faddeev differential equations. Using this approach allows one to overcome
technical complications due to the strong repulsive part in the atomic
interaction. Along the same line we now study the asymmetric $^3$He$^4$He$_2$
system. We calculate its bound states, the scattering phase shifts
of a $^3$He atom off a $^4$He dimer at ultra-low energies, and
the corresponding scattering lengths.

In the present investigations we employ two realistic He-He atomic
interactions, the LM2M2 potential by Aziz and Slaman \cite{Aziz91} and the
TTY potential by Tang, Toennies and Yiu \cite{Tang95}. As in \cite{KMS-JPB},
\cite{MSSK} we use the finite-difference approximation of the two-dimensional
partial-wave Faddeev equations.  We consider the case of zero total angular
momentum and take $\hbar^2/m_{^4{\rm He}}=12.12$\,K\,\AA$^2$ where $m_{^4{\rm
He}}$ stands for the mass of a ${^4{\rm He}}$ atom. Notice that the
$^4$He-dimer energy is 1.30348\,mK for the LM2M2 potential and 1.30962\,mK
for the TTY potential \cite{MSSK}.
\begin{table}[h]
\caption{\small Absolute value of the $^3$He$^4$He$_2$ trimer binding
energy (in mK). Grids used:  $N_\rho = 600$, $N_\theta = 605$, the cut-off
hyperradius $\rho_{\rm max} = 200$\,{\AA}.}
\label{tableTrimerGS}
\centering
\begin{tabular}{|r|rrrrrr|}
\hline
\hline
Potential & \quad $l_{\rm max}$ &  This work & \quad\cite{EsryLinGreene}&
\quad \cite{Nielsen}
& \quad \cite{Bressani} & \quad\cite{Roudnev} \\
\hline
\hline
       & 0 &  7.30 & \quad 10.22 &       & & \\
        \cline{2-3}
LM2M2  & 2 & 13.15 &       &       & & \\
        \cline{2-3}
       & 4 & 13.84 &       & \quad 13.66 &        & \quad 14.4\\
\hline
\hline
      & 0 &  7.25 &       &       & & \\
        \cline{2-3}
TTY    & 2 & 13.09 &       &       & & \\
        \cline{2-3}
       & 4 & 13.78 &       &       & \quad 14.165 & 14.1 \\
\hline
\hline
\end{tabular}
\end{table}

Due to the smaller mass of the $^3$He atom ($m_{^3{\rm
He}}/m_{^4{\rm He}}$=$0.753517$), the $^3$He -- $^4$He
system is unbound. None the less, the $^3$He$^4$He$_2$
trimer exists, though with a binding energy one order
of magnitude smaller than the one of the $^4$He$_3$ ground state
(see, e.g. \cite{KMS-JPB} and \cite{MSSK}). And, in contrast to the
symmetric case, there is no excited (Efimov-type) state in the
asymmetric $^3$He$^4$He$_2$ system.

\begin{figure}[h]
\vspace*{-1.5cm}
\includegraphics[angle=-90,width=12.cm]{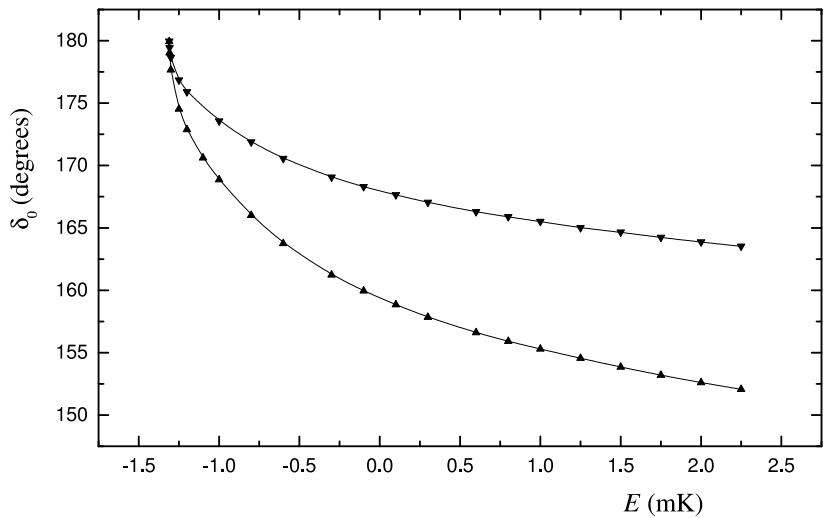}
\vspace*{-0.5cm}

\caption{\small $S$-wave $^3$He atom -- $^4$He dimer scattering phase
shifts $\delta_0(E)$ in case of the TTY He--He interatomic potential as a
function of the c.m. energy. The lower curve corresponds to the case where
$l_{\rm max}=0$ while for the upper curve $l_{\rm max}=2$. The corresponding
phase shift curves for the LM2M2 potential are practically the same.}
\label{Fig-phases}
\end{figure}

The best dimensions of the grids employed in our investigation were $N_\rho =
600$ and $N_\theta = 605$ with a cut-off hyperradius $\rho_{\rm max} =
200$\,{\AA} (see \cite{KMS-JPB} for explanation of the notation). The maximal
value of the angular momenta taken into account in the two-body subsystems
was $l_{\rm max}=4$. Our results for the $^3$He$^4$He$_2$ binding energy as
well as the results available in the literature, are presented in Table
\ref{tableTrimerGS}. One can see that at this point the different approaches
agree fairly well. A small discrepancy in the results is possibly due to
limitations of the computer facilities. We have found that most of the
contribution to the $^3$He$^4$He$_2$ binding energy stems from the $l=0$ and
$l=1,2$ partial waves, about 53\% and 42\%, respectively. The overall
contribution from the $l=3$ and $l=4$ components is of the order of 5\,\%,
that is, approximately the same as in the case of the symmetric $^4$He trimer
\cite{MSSK}. A certain (but rather small) deepening of the $^3$He$^4$He$_2$
binding energy  may be expected when choosing a larger grid.

Our phase shift results obtained for a grid with $N_\rho=502$,
$N_\theta=500$, and $\rho_{\rm max}=460$\,{\AA} are plotted in
Fig.~\ref{Fig-phases}. Note that exactly the same grid was employed in the
$^4$He--$^4$He$_2$ phase-shift calculations of \cite{MSSK}. Incident
energies below and above the breakup threshold, i.\,e. both for the elastic
scattering $(2+1\longrightarrow 2+1)$ and breakup $(2+1\longrightarrow
1+1+1)$ processes, were considered. Table \ref{tableTrimerLen}
contains our results for the $^3$He--$^4$He$_2$ scattering length.

\begin{table}[h]
\caption{\small Estimations for the $^{3}$He atom\,--\,$^4$He dimer
scattering length (in \AA). Grids used are the same as in Table
\ref{tableTrimerGS}.} \label{tableTrimerLen}

\centering
\begin{tabular}{|c|cc|cc|}
\hline\hline
& & & & \\[-2ex]
 & \multicolumn{2}{c|}{TTY} & \multicolumn{2}{c|}{LM2M2} \\
\cline{2-5}
$\qquad l_{\rm max}\qquad$ & $\quad$This work$\quad$
& $\quad$\cite{Roudnev}$\quad$ & $\quad$This work$\quad$
& $\quad$\cite{Roudnev}$\quad$ \\
\hline\hline
& & & &  \\[-2ex]
0  & $38.8$ &  &38.5 & \\
2  & $22.4$ &  & 22.2&   \\
4  & $21.2$  &  19.6 &21.0 & 19.3 \\
\hline\hline
\end{tabular}
\end{table}

\noindent{\small {\bf Acknowledgements.} We are grateful to
Prof.~V.\,B.\,Belyaev and Prof.~H.\,Toki for providing us with the
possibility to perform calculations at the supercomputer of the Research
Center for Nuclear Physics of Osaka University, Japan.}


\begin{thebibliography}{99}

\bibitem{GrebToeVil}  S. Grebenev, J.P. Toennies, and A.F.
Vilesov: Science {\bf 279} (1998) 2083.

\bibitem{DimerExp1} F. Luo, C. F. Giese, and W. R. Gentry:
       J. Chem. Phys. {\bf 104} (1996) 1151.

\bibitem{CGM} J. Carbonell, C. Gignoux, and S.P. Merkuriev:
     Few--Body Systems {\bf 15} (1993) 15.

\bibitem{RoudnevYakovlev} V. Roudnev and S. Yakovlev:
        Chem. Phys. Lett. {\bf 328} (2000) 97 (arXiv:\,physics/9910030).

\bibitem{KMS-JPB} E.A. Kolganova, A.K. Motovilov, and S.A. Sofianos:
J.~Phys. B {\bf 31} (1998) 1279 (arXiv:\,physics/9612012).

\bibitem{MSSK} A.K. Motovilov, W. Sandhas, S.A. Sofianos, and E.A.
Kolganova: Eur.~Phys.~J. D {\bf 13} (2001) 33 (arXiv:\,physics/9910016).

\bibitem{Barletta} P. Barletta and A. Kievsky: Phys. Rev. A
  {\bf 64} (2001) 042514.

\bibitem{EsryLinGreene} B.D.~Esry, C.D.~Lin, and C.H.~Greene:
       Phys. Rev.~A {\bf 54} (1996) 394.

\bibitem{Nielsen} E. Nielsen, D.V. Fedorov, and A.S. Jensen:
  J. Phys. B {\bf 31} (1998) 4085 (ar\-X\-iv: physics/9806020).

\bibitem{Bressani} D. Bressanini, M. Zavaglia, M. Mella, and G.
Morosi: J. Chem. Phys. {\bf 112} (2000) 717.

\bibitem{Roudnev} V. Roudnev: Talk at the 17th International
IUPAP Conference on Few-Body Problems in Physics (Durham, North Carolina,
USA, June 5--10, 2003).

\bibitem{Aziz91} R.A. Aziz and M.J. Slaman: J. Chem. Phys. {\bf 94}
      (1991) 8047.

\bibitem{Tang95} K.T. Tang, J.P. Toennies, and C.L. Yiu:
     Phys. Rev. Lett. {\bf 74} (1995) 1546.

\end{thebibliography}
\end{document}